**Obituary: Dr. Dianne Johnson (1947 - 2012)**

It is with great sadness that I report the passing of Dr Dianne Johnson in May 2012. Dr Johnson was a pivotal and important figure in the field of Australian cultural astronomy and in the campaign for Aboriginal rights.

Dianne Dorothy Johnson was born on 11 January 1947, attended Newcastle Girls High, and enrolled in the Sydney Kindergarten Teachers College to study early childhood education. She left Australia in 1968 to take a position as a preschool director in Port Moresby, Papua New Guinea. Her cultural experience in Papua inspired her to pursue a degree in anthropology at the University of Sydney, which she completed with First Class Honors in 1971. She then completed a PhD at the University of Sydney in 1984 studying the role of powerful women in the PNG government after it gained independence from Australia in 1975. Dr Johnson lectured in anthropology at the University of Sydney and at the Sydney Kindergarten Teachers College before moving to the Blue Mountains (west of Sydney) in 1988 to become director of the Katoomba-Leura Preschool.

Throughout her career, she was active in Indigenous affairs and worked closely local with Aboriginal communities and elders. She was a prolific writer, authoring books such as *The Jewel Box* (1990), *Mandatory Injustice* (2000)*, Lighting the Way* (2002), *Aunty Joan Cooper* (2003), *Sacred Waters* (2007), *Hut in the Wild* (2011), and *Bruny d'Entrecasteaux and his encounter with Tasmanian Aborigines* (2012)*. Sacred Waters* won Johnson the 2008 History Prize from the Premier of New South Wales.

Dr Johnson is perhaps best known in the astronomy community for her work on Australian cultural astronomy. In 1986, the arrival of Comet Halley sparked her interest in the night sky, prompting her to write a book on the astronomical traditions of Indigenous Australians. While working on her manuscript, she decided to immerse herself in an astrophysics course at the University of Sydney. Her husband, George Zdenkowski, said the subject interested her so much that she casually considered taking a post-doctoral position in astrophysics. Her book, *Night Skies of Aboriginal Australia: a Noctuary* (1998) remains the most comprehensive overview on Australian cultural astronomy to date, and is now in its second edition.

In her noctuary, she explores the astronomies of Indigenous Australians. Dr Johnson advocated the use of the term "astronomies" over the singular term "astronomy" to highlight the varied and distinct astronomical traditions of the hundreds of Indigenous cultures across Australia. Using an anthropological approach, her book discusses the influence of the sun, moon, and stars on topics such as cosmology, natural cycles, mythology, history, social relations, kinship ties, healing, and astronomical observations. Dr Johnson called her book a noctuary to emphasize that this was a record of culture at night; a nocturnal journal for learning laws, customs, and traditions.

Dr Johnson was fascinated with the beauty of Indigenous sky stories, particularly those that showed commonality among diverse groups of people. Of particular interest to her were stories about the Pleiades star cluster – referred to as the Seven Sisters throughout much of Aboriginal Australia. This was the topic on which Dr Johnson spoke at both the *Ilgarijiri – Things Belonging to the Sky* symposium at the Australian Institute for Aboriginal and Torres Strait Islander Studies (Canberra) in 2009 and the International Society for Archaeoastronomy and Astronomy in Culture conference in Lima, Peru in 2011. It was at the AIATSIS meeting in 2009, during the International Year of Astronomy, that I first met Dr Johnson. She was genuine and friendly and we talked extensively about the influence of her book and the future of cultural astronomy in Australia. When I saw her again in Peru, she spoke passionately about her desire to establish night sky preserves in Australia. This was not only to preserve dark skies for astrophysical research, but also to save our collective astronomical heritage – both Indigenous and colonial.

Dr Johnson passed away on 3 May 2012. She leaves behind a long and proud legacy. She will be honored and remembered for her contribution to Australian cultural astronomy, her campaign for Indigenous rights, and her passion for education.

*Duane W. Hamacher*
*University of New South Wales*
*Sydney, Australia*

*I would like to acknowledge Malcolm Brown and George Zdenkowski for information about Dr Johnson's early life. Further details of her biography can be found here:*
*http://www.smh.com.au/comment/obituaries/anthropologist-who-wrote-to-seek-justice-20120601-1zmyw.html*

**Published Works by Dianne Johnson**

Johnson, D.D., 1984. From fairy to witch: imagery and myth in the Azaria case. *Australian Journal of Cultural Studies*, 2(2), 90-106.

Johnson, D.D., 1984. *The Government Women: gender and structural contradiction in Papua New Guinea*. PhD Thesis, Department of Anthropology, University of Sydney.

Johnson, D.D., 1990. *The Jewel Box*. (Illustrated by R. Purdon.) Mumbai, India, Butterfly Books.

Johnson, D.D., 1998. *Night Skies of Aboriginal Australia: a Noctuary*. Sydney, University of Sydney Press.

Johnson, D. 2000, *The Pleiades in Australian Aboriginal and Torres Strait Islander Astronomies*. In S. Kleinert, M. Neale, and R. Bancroft (eds) *The Oxford companion to Aboriginal art and culture*. Melbourne, Oxford University Press. Pp. 22-24.

Johnson, D.D., and Zdenkowski, G., 2000. *Mandatory injustice: compulsory imprisonment in the Northern Territory*. Sydney, Australian Centre for Independent Journalism, University of Technology, Sydney.

Johnson, D.D., 2002. *Lighting the way: reconciliation stories*. Sydney, The Federation Press.